\documentclass[reprint,aps,prd,amsmath,amssymb,floatfix,nofootinbib]{revtex4-2}
\usepackage[utf8]{inputenc} 
\usepackage[T1]{fontenc}    
\usepackage{amsmath, amssymb, bm, slashed, physics}
\usepackage{graphicx}
\usepackage{caption}
\usepackage{subcaption} 
\usepackage{float}      
\usepackage{multirow}
\usepackage{makecell}
\usepackage{enumerate}
\usepackage{comment}
\usepackage{xcolor}
\usepackage{hyperref}
\hypersetup{
    colorlinks=true,
    linkcolor=blue,
    citecolor=blue,
    urlcolor=blue
}
\usepackage{url}
\usepackage{silence}
\hypersetup{
    colorlinks=true,
    citecolor=blue,
    linkcolor=blue,
    urlcolor=blue
}
\begin{document}
\title{\texorpdfstring{Probing the dimuon channel of a $Z'$ boson at the HL-LHC using multivariate analysis}{Probing the dimuon channel of a Z' boson at the HL-LHC using multivariate analysis}}
\author{Ali Muhammad H. H.}
\email{ali.hamed@bue.edu.eg}
\affiliation{Physics Department, Faculty of Science, Ain Shams University, Cairo, Egypt}
\affiliation{Basic Science Department, Faculty of Engineering, The British University in Egypt, P.O. Box 43, El Sherouk City, Cairo 11837, Egypt}

\author{El-sayed A. El-dahshan}
\affiliation{Physics Department, Faculty of Science, Ain Shams University, Cairo, Egypt}

\author{S. Elgammal}
\affiliation{Centre for Theoretical Physics, The British University in Egypt, P.O. Box 43, El Sherouk City, Cairo 11837, Egypt}
\begin{abstract}
The upcoming upgrade of the existing LHC facility at CERN is known as the High-Luminosity Large Hadron Collider (HL-LHC). It is designed to extend its physics reach by substantially increasing the integrated luminosity. It will enable more precise measurements of the Standard Model (SM) and improve sensitivity to rare events and possible new physics signatures. This study adopts a multivariate analysis (MVA) approach to effectively discriminate the dark Higgs (DH) signal against the dominant SM background. The analysis targets the leptonic decay mode of the $Z'$ boson, focusing on the dimuon final state at $\sqrt{s} = 14,\text{TeV}$ and $3000,\text{fb}^{-1}$ integrated luminosity corresponding to the HL-LHC. The DH signal is examined using the Toolkit for Multivariate Analysis (TMVA), employing and comparing the performance of three classifiers: Boosted Decision Trees (BDT), Deep Neural Networks (DNN), and Likelihood estimators.
\begin{description}
\item[Keywords]
Dark Matter, Dark Higgs, Multivariate Analysis, The HL-LHC
\end{description}
\end{abstract}
\maketitle
\section{Introduction}
\label{sec:intro}
Dark matter (DM) is a hypothetical form of matter that does not interact electromagnetically, making it invisible to all current forms of electromagnetic observation~\cite{Bertone2005,Feng2010}. Evidence for its presence arises from a variety of astrophysical and cosmological measurements, such as galaxy rotation curves~\cite{Rubin1980,Sofue2001}, gravitational lensing phenomena~\cite{Clowe2006,Kneib2011}, and fluctuations observed in the cosmic microwave background~\cite{Planck2018,WMAP9}.
These observations collectively suggest that DM constitutes about 27\% of the total energy density of the universe~\cite{Planck2018,PDG2022}.

Despite its gravitational influence, its fundamental particle properties remain elusive. Many potential candidates have been suggested, such Weakly Interacting Massive Particles (WIMPs)~\cite{Jungman1996,Arcadi2018, Mahmoud2024}, axions~\cite{Peccei1977,Sikivie1983}, sterile neutrinos~\cite{Dodelson1994,Abazajian2001}, and particles from dark sectors such as dark photons or $Z'$ bosons~\cite{Alexander2016,Holdom1986}. Various experimental strategies are employed to search for DM, including direct detection through nuclear recoil measurements~\cite{Aprile2018,Cui2017}, indirect searches based on astrophysical observations~\cite{Gaskins2016}, and collider-based approaches targeting missing transverse energy signatures~\cite{Boveia2018,Aaboud2018}.

Although the Standard Model (SM) has achieved great success, most recently with the discovery of the Higgs boson by the ATLAS and CMS collaborations in 2012~\cite{CMSHiggs2012,ATLAS_Higgs2012}, it nevertheless fails to resolve fundamental issues, such as uncovering the particle nature of DM. Therefore, physicists have explored beyond‐the‐Standard‐Model (BSM) scenarios. One such framework is the Dark Higgs (DH) model with a mono-$Z'$ portal~\cite{MonoZ}. In this framework, DM particles are generated in proton–proton (pp) collisions via a heavy neutral mediator, the $Z'$ boson. This scenario has been studied in both Run-I~\cite{Elgammal2023} and Run-II~\cite{ATLAS2023} of the LHC.

This work examines the DH model at the High-Luminosity phase of the LHC (HL-LHC), which is the upcoming upgrade of the LHC, operating at $14\,\mathrm{TeV}$ center-of-mass energy ($\sqrt{s}$), and a high integrated luminosity ($\mathcal{L}$) corresponding to $3000\,\mathrm{fb}^{-1}$ ~\cite{HL-LHC-WhitePaper,HL-LHC-TDR}. Signal and SM background samples are generated using Monte Carlo (MC) simulations.

To distinguish the signal from the dominant SM backgrounds, the work employs multivariate analysis (MVA) using the TMVA package built in the ROOT framework. This approach has been applied in previous works~\cite{CMS:2025muon,Choudhury2024,Bhardwaj2020Boosted,Rana2024Analytical} and in recent HL-LHC studies~\cite{Baradia2024,Chang2020Trilinear,Adhikary2020Prospects}. We compare three classifiers: Boosted Decision Trees (BDT), Deep Neural Networks (DNN), and Likelihood estimators, which were used in previous studies, such as~\cite{Baradia2024,Chang2020Trilinear,Adhikary2020Prospects, Roe:2005,Baldi:2014kfa,Cowan:2010js,S0022286022024802}.   

The structure of the paper is as follows. Section~\ref{section:model} briefly introduces the simplified DH model. Section~\ref{section:HL-LHC} describes the HL-LHC. Section~\ref{section:signal&background} presents the techniques used for event generation in the MC simulations of the DH signals and its corresponding SM backgrounds. The TMVA package is introduced in Section~\ref{section:TMVA}, and the MVA and its results are discussed in Section~\ref{section:Analysis}. Finally, Section~\ref{section:Summary} provides a summary and conclusion.
\section{The Simplified Model in the Mono-\texorpdfstring{\boldmath$Z'$}{Z'} Portal}
\label{section:model}

\begin{figure}[htbp]
  \centering
  \includegraphics[width=0.3\textwidth]{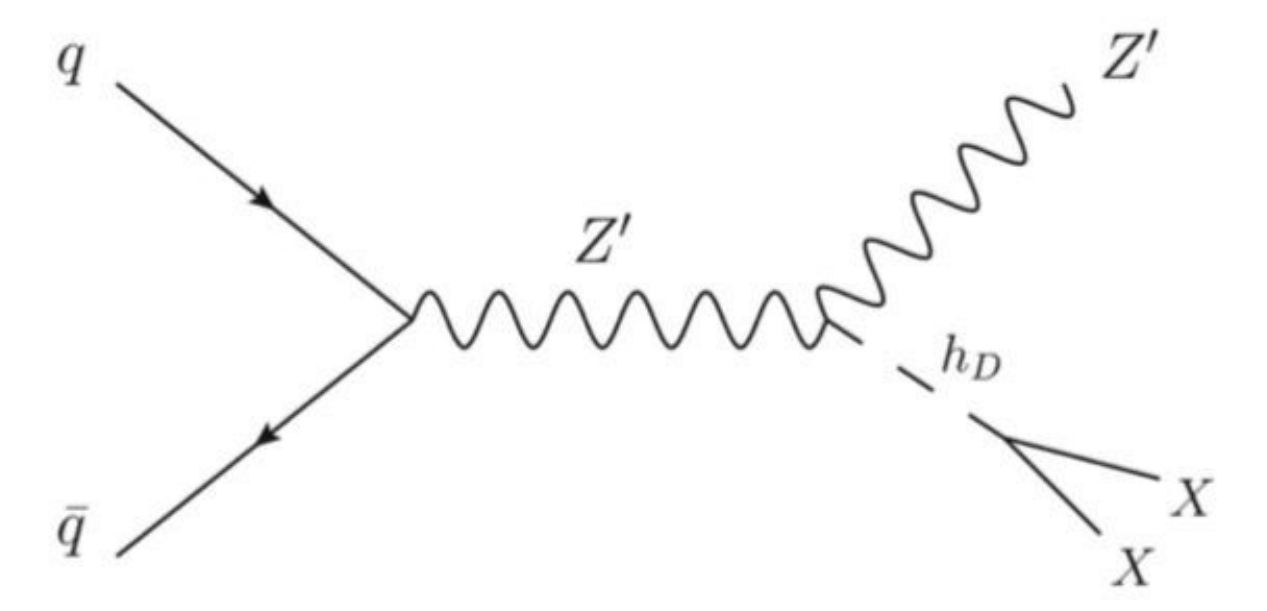} 
  \caption{Feynman diagram of the DH model, adapted from Ref.~\cite{MonoZ}}
  \label{fig:Feynmann} 
\end{figure}

The study in Ref.~\cite{MonoZ} proposed the potential generation of DM particles associated with a neutral heavy gauge boson, denoted as $Z'$, through the fusion of $q\bar{q}$ pairs in pp collisions. In this framework, the $Z'$ mediates the interaction between SM particles and dark sector particles, with coupling constants $g_{SM}$ and $g_{DM}$, respectively. Three simplified models were introduced in~\cite{MonoZ}: the light vector (LV), the LV with inelastic effective field theory (EFT) coupling, and the DH models, with the DH model being the main focus of this paper. In the DH model, the $Z'$ boson mediates the production of a dark-sector Higgs boson ($h_D$), which subsequently decays into a pair of DM particles ($\chi\bar{\chi}$), following the heavy dark-sector (HDS) mass assumptions summarized in Table~\ref{tab:mass_assumptions}. The Feynman diagram illustrating the DH model is shown in Fig.~\ref {fig:Feynmann}.

This study adopted the HDS mass assumption where $M_{Z'} = M_{h_D}$ for $M_{Z'} > 125\,\mathrm{GeV}$ and takes the muonic decay of the $Z'$, into dimuon, plus $E_T^{\mathrm{miss}}$, referring to DM particles, as the final signature of the interaction. The free parameters of the model are $M_{Z'}$, $M_{h_D}$, $g_{SM}$ and $g_{DM}$. The scope of this study for $M_{Z'}$, and $M_{h_D}$ ranges from 200 to 1000 $\mathrm{GeV}$, choosing $g_{SM} = 0.25$ and $g_{DM} = 1.0$ as recommended by~\cite{ATLAS2023} to get the highest cross sections multiplied by the branching ratios ($\sigma \times \mathrm{Br}$) for the signal. Also, the range of 0.02 to 0.2 for $g_{SM}$ was excluded for mass ranges from 200 to 1000 $\mathrm{GeV}$ by that study~\cite{ATLAS2023} performed by ATLAS. The studies carried out in Ref.~\cite{MonoZ,Elgammal2023} found that $\sigma\times\mathrm{Br}$ does not depend on the choice of the DM mass but on $M_{Z'}$ and $M_{h_D}$.

This work aims to probe the DH model at the HL-LHC, corresponding to $\sqrt{s} = 14,\text{TeV}$ and $\mathcal{L} = 3000,\text{fb}^{-1}$, using MC simulated samples generated as described in Section~\ref{section:signal&background} with $g_{SM} = 0.25$ and $g_{DM} = 1.0$ and $M_{Z'} = M_{h_D} \in [200, 1000]\,\mathrm{GeV}$.
\begin{table}[htbp]
\centering
\begin{tabular}{|l | l|}
\hline
\textbf{Scenario} & \textbf{Mass assumptions} \\
\hline
Heavy dark sector & $M_{h_D} = \begin{cases}
                   125\,\mathrm{GeV}, & M_{Z'} < 125\,\mathrm{GeV},\\
                   M_{Z'},            & M_{Z'} > 125\,\mathrm{GeV}.
\end{cases}$
                  \\
\hline
\end{tabular}
\caption{Mass assumptions of the HDS for the particles produced in the DH model as outlined in Ref.~\cite{MonoZ}.
}
\label{tab:mass_assumptions}
\end{table}
\section{The HL-LHC}
\label{section:HL-LHC}
The upcoming major upgrade to the current LHC at CERN is the HL-LHC, designed to extend its physics reach by substantially increasing $\mathcal{L}$. With a target of delivering up to $3000\,\mathrm{fb}^{-1}$ of pp collision data at $\sqrt{s} = 14\,\mathrm{TeV}$, the HL-LHC will enable precision studies of the SM and enhance the sensitivity to rare processes and potential signals of new physics BSM~\cite{HL-LHC-WhitePaper,HL-LHC-TDR}.

This upgrade involves the implementation of advanced superconducting magnets, new cryogenic systems, and improved beam collimation to accommodate higher beam intensities. The ATLAS and CMS experiments will also be significantly upgraded to handle the increased data rates and radiation levels expected under high-luminosity conditions~\cite{ATLAS-Upgrade,CMS-Upgrade}. The HL-LHC program represents a critical step toward probing the deepest questions in particle physics, offering unprecedented discovery potential in the search for phenomena such as DM mediators, long-lived particles, and extended Higgs sectors. In this study, the produced events are MC simulations at the HL-LHC.
\section{Generation of the Monte Carlo Simulations}
\label{section:signal&background}
\subsection{Generation of DH Signal Samples}
The MC simulation samples of the DH signal were simulated employing 
\mbox{\texttt{MadGraph5\_aMC@NLO}}\allowbreak~v3.5.0~\cite{Madgraph} at next-to-leading order (NLO). The events were then combined with \mbox{\texttt{Pythia8}}\allowbreak~\cite{Pythia} for hadronization and parton showering, at $\sqrt{s} = 14\,\mathrm{TeV}$. The CMS detector response was simulated using \texttt{Delphes}~\cite{Delphes}, employing the CMS Pile-Up configuration card. The produced events were generated for the HDS masses assumption listed in tabled~\ref{tab:mass_assumptions} for $M_{Z'} \in [200, 1000]$ taking $g_{SM} = 0.25$, and $g_{DM} = 1.0$ as chosen in~\cite{ATLAS2018,ATLAS2023,Elgammal2025}.

For the $Z'$ decaying into dimuons, $\sigma \times \mathrm{Br}(Z' \to \mu\mu)$ denotes the production cross section multiplied by the branching ratio. The values for all benchmark points (BPs) of the DH signal samples are listed in Table~\ref{tab:cross_section_table}.
\begin{table}[htbp]
\centering
\begin{tabular}{|c|c|c|}
\hline
\textbf{BPs} & $\bm{M_{Z'}\,\mathrm{(GeV)}}$ & $\bm{\sigma\times\mathrm{Br}(Z'\rightarrow\mu\mu)\,\mathrm{(pb)}}$ \\
\hline
BP1 & 200 & $6.23 \times 10^{-2}$ \\
BP2 & 300 & $1.38 \times 10^{-2}$ \\
BP3 & 400 & $3.93 \times 10^{-3}$ \\
BP4 & 500 & $1.48 \times 10^{-3}$ \\
BP5 & 600 & $6.46 \times 10^{-4}$ \\
BP6 & 700 & $3.09 \times 10^{-4}$ \\
BP7 & 800 & $1.58 \times 10^{-4}$ \\
BP8 & 900 & $8.50 \times 10^{-5}$ \\
BP9 & 1000 & $4.75 \times 10^{-5}$ \\
\hline
\end{tabular}
\caption{All BPs considered in the DH model along with their $\sigma\times\mathrm{Br}(Z'\rightarrow\mu\mu)$ in Pb for $M_{Z'}$ ranging from 200 to 1000 $\mathrm{GeV}$, taking $g_{SM} = 0.25$ and $g_{DM} = 1.0$. All cross sections are computed employing \texttt{MadGraph5\_aMC@NLO}~v3.5.0~\cite{Madgraph} at NLO, combined with \texttt{Pythia8}~\cite{Pythia} for hadronization and parton showering, at $\sqrt{s} = 14\,\mathrm{TeV}$.}
\label{tab:cross_section_table}
\end{table}
\subsection{Generation of SM Background Samples}
As this study is concerned with events that have a final signature of the form $\mu^{+}\mu^{-} + E^{miss}_{T}$, the background processes of the SM produced are those listed in Table~\ref{tab:SM_backgrounds} where $E_T^{\mathrm{miss}}$ in their topology refers to undetected neutrinos. All these processes are simulated employing \texttt{MadGraph5\_aMC@NLO}~v3.5.0~\cite{Madgraph} at NLO. The events were then combined with \texttt{Pythia8}~\cite{Pythia} for hadronization and parton showering, at $\sqrt{s} = 14\,\mathrm{TeV}$. The SM backgrounds along with their corresponding $\sigma\times\mathrm{Br}$ are listed in Table~\ref{tab:SM_backgrounds}, as well. The CMS detector response for the SM backgrounds was also simulated using \text{Delphes} ~\cite{Delphes}, employing the CMS Pile-Up configuration card, as was done for the DH signals. 

The discrimination of the signal from the overwhelming background is always challenging. Hence, this study employs the machine learning methods available in the toolkit for MVA known as TMVA.
\begin{table}[htbp]
\centering
\begin{tabular}{|c|c|}
\hline
\textbf{Process} & $\bm{\sigma\times\mathrm{Br}\ \mathrm{(pb)}}$ \\
\hline
Drell-Yan ($DY$) & 2176.76 \\
$t\bar{t}$ & 6.72 \\
Single top ($\bar{t}W^+$, $tW^-$) & $32.25 \times 10^{-2}$ \\
$W^+W^-$ & $84.97 \times 10^{-2}$\\
$W^\pm Z$ & $9.96 \times 10^{-2}$ \\
$ZZ \to 2\mu2\nu$ & $4.60 \times 10^{-2}$ \\
$ZZ \to 4\mu$ & $1.16 \times 10^{-2}$ \\
\hline
\end{tabular}
\caption{SM background processes that resemble the DH signal are summarized, along with their $\bm{\sigma\times\mathrm{Br}}$ in pb computed employing \texttt{MadGraph5\_aMC@NLO}~v3.5.0~\cite{Madgraph} at NLO, combined with \texttt{Pythia8}~\cite{Pythia}, at $\sqrt{s} = 14\,\mathrm{TeV}$.}
\label{tab:SM_backgrounds}
\end{table}
\section{TMVA: Toolkit for Multivariate Analysis}
\label{section:TMVA}
The TMVA package is an integral component of the ROOT data analysis framework, providing a broad range of machine learning algorithms for classification and regression tasks in high-energy physics (HEP)~\cite{TMVA,Roe:2005}. TMVA is widely adopted in particle physics analyses due to its seamless integration with ROOT ~\cite{ROOT1997}, support for ROOT I/O, and a comprehensive suite of pre-processing, training, and evaluation tools. Among the available classifiers are BDT, DNN, and Likelihood estimators, making it a versatile platform for complex pattern recognition problems encountered in collider experiments.

TMVA allows users to train, test, and evaluate models with minimal overhead while providing detailed diagnostics such as variable ranking, receiver operating characteristic (ROC) curves, and overtraining checks~\cite{TMVA,Choudhury2024}. Its flexibility in defining custom input variables and optimization strategies has made it a popular choice in LHC analyses and phenomenological studies~\cite{Therhaag2011}. In this work, TMVA is utilized to implement and compare the performance of the three classifiers, BDT, DNN, and Likelihood estimators, in distinguishing the DH signal from the SM backgrounds.
\section{Analysis and Results}
\label{section:Analysis}
After generating the simulated events for the signals and backgrounds, it is time for the reconstruction of the final states containing two high-$p_T$ muons with opposite charges, accompanied by significant $E_T^{\mathrm{miss}}$ arising from DM candidates. The pre-selection criteria applied to muons are summarized in Table~\ref{table:Pre_Selection}.

Muons are selected by requiring a transverse momentum $p_T^\mu > 30~\mathrm{GeV}$ and a pseudorapidity within $|\eta^\mu| < 2.4$. To ensure good isolation and exclude contamination from hadronic jets, they must meet the "IsolationVarRhoCorr" condition as defined in \texttt{Delphes}~\cite{Delphes}, which corrects for pileup. This criterion ensures that the sum of transverse momenta within a $\Delta R = 0.5$ cone (excluding the muon) does not exceed 10\% of the $p_T^\mu$. Additionally, the invariant mass of the dimuon system ($M_{\mu\mu}$) must be greater than $60~\mathrm{GeV}$ to focus on heavy resonance searches.
\begin{table}[htbp]
  \centering
  \begin{tabular}{|c|c|c|}
    \hline
    \textbf{Step} & \textbf{Criteria} & \textbf{Requirements} \\ [1.0ex]
    \hline
    \multirow{4}{*}{Pre-selection} 
      & $p_{T}^{\mu}$ (GeV) & $> 30$ \\
      & $|\eta^{\mu}|$ (rad) & $< 2.4$ \\
      & IsolationVarRhoCorr & $< 0.1$ \\
      & $M_{\mu^{+}\mu^{-}}$ (GeV) & $> 60$ \\
    \hline
  \end{tabular}
  \caption{
    Overview of the pre-selection criteria used for the event selection before the MVA.
  }
  \label{table:Pre_Selection}
\end{table}
\begin{figure}[htbp]
  \centering
  \includegraphics[width=0.4\textwidth]{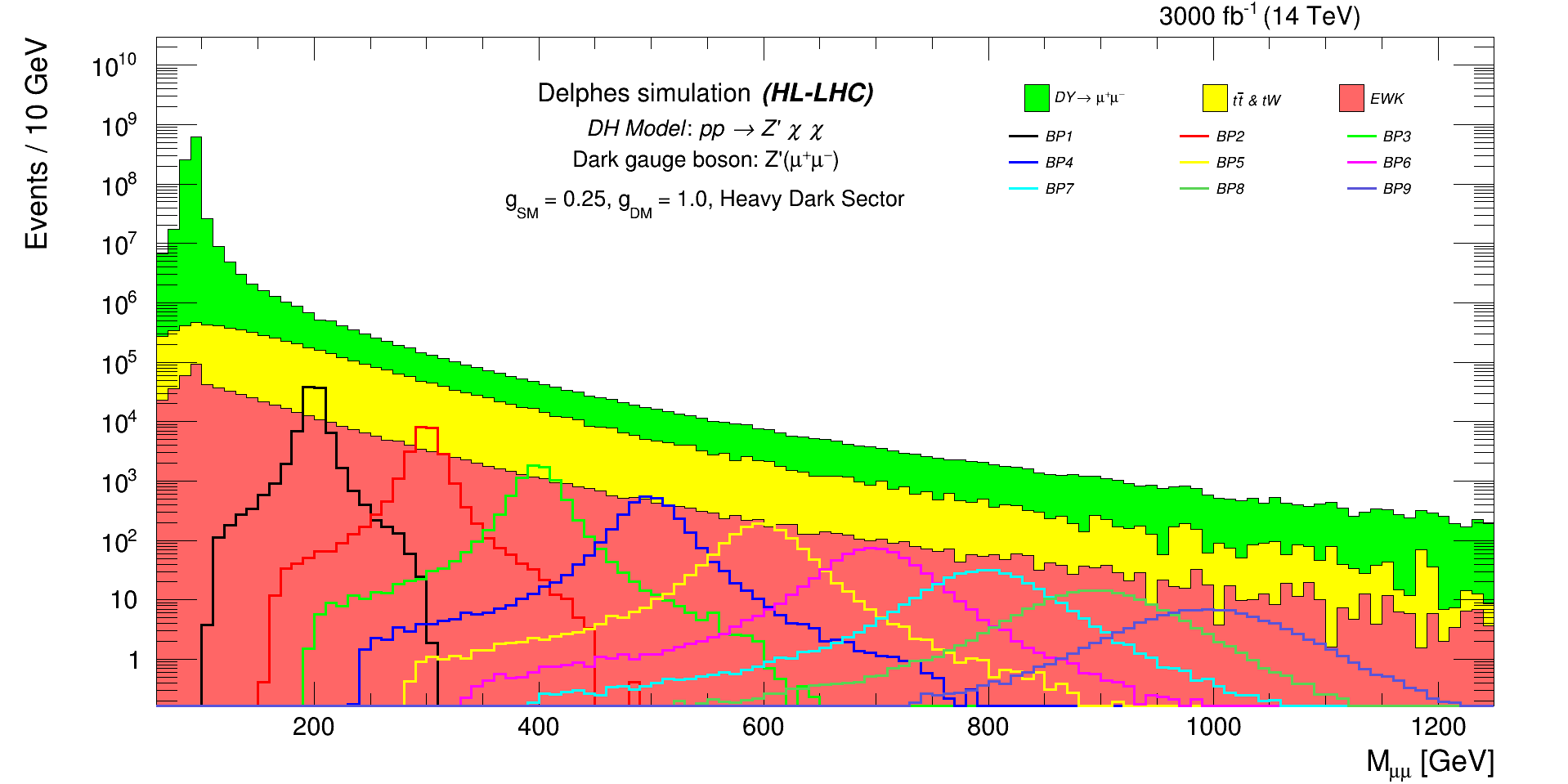} 
  \caption{Distribution of dimuon invariant mass $M_{\mu\mu}$ for DH signals, with $M_{Z'}$ ranging from 200 to 1000 $\mathrm{GeV}$ and taking $g_{SM} = 0.25$ and $g_{DM} = 1.0$, is overlaid on its corresponding SM background summarized in Table~\ref{tab:SM_backgrounds} following the pre-selection criteria listed in Table~\ref{table:Pre_Selection}, with $\mathcal{L} = 3000,\mathrm{fb}^{-1}$ and at $\sqrt{s} = 14,\mathrm{TeV}$.}
  \label{fig:Mass_Dist} 
\end{figure}
\begin{table*}[htbp]
    \centering
    \begin{tabular}{|c|c|}
    \hline
    Variable & Description \\
    \hline
    $P_{T}^{\mu_{1}}$ & Transverse momentum of the first muon\\
    $P_{T}^{\mu_{2}}$ & Transverse momentum of the second muon \\
    $E_T^{\mathrm{miss}}$ & Missing transverse energy \\
    $M_{\mu\mu}$ &  Invariant mass of dimuon\\
    $\Delta\Phi$ & The azimuthal angle between the directions of the dimuon and the $E_{T}^{\mathrm{miss}}$ \\
    \hline
    \end{tabular}
    \caption{List of kinematic variables employed in this MVA.}
    \label{tab:Input_Vars_table}
\end{table*}
\begin{figure*}[htbp]
  \centering
      \begin{subfigure}[b]{0.4\textwidth}
    \includegraphics[width=\textwidth]{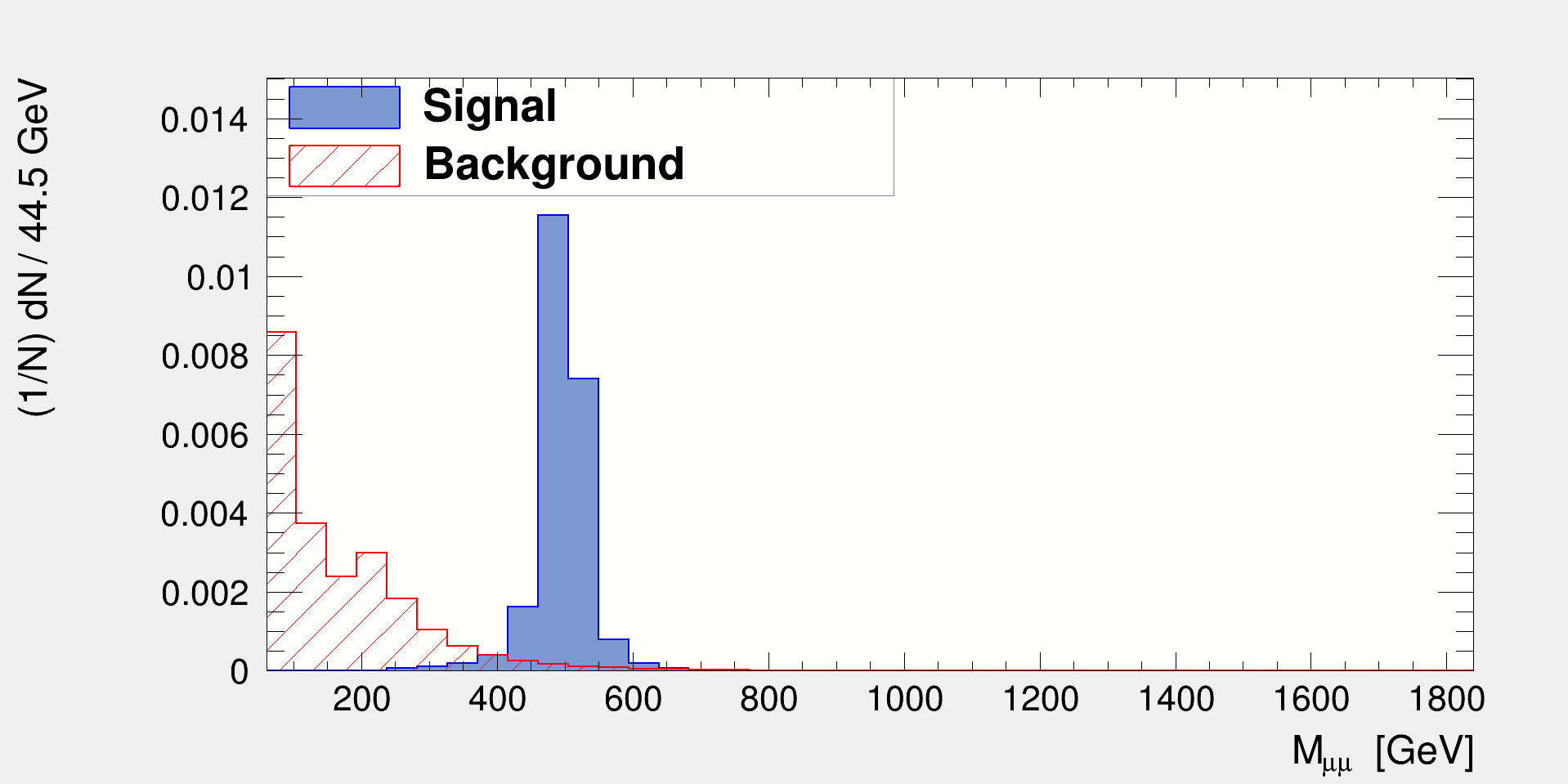}
    \caption{Invariant mass of dimuon}
  \end{subfigure}
    \begin{subfigure}[b]{0.4\textwidth}
    \includegraphics[width=\textwidth]{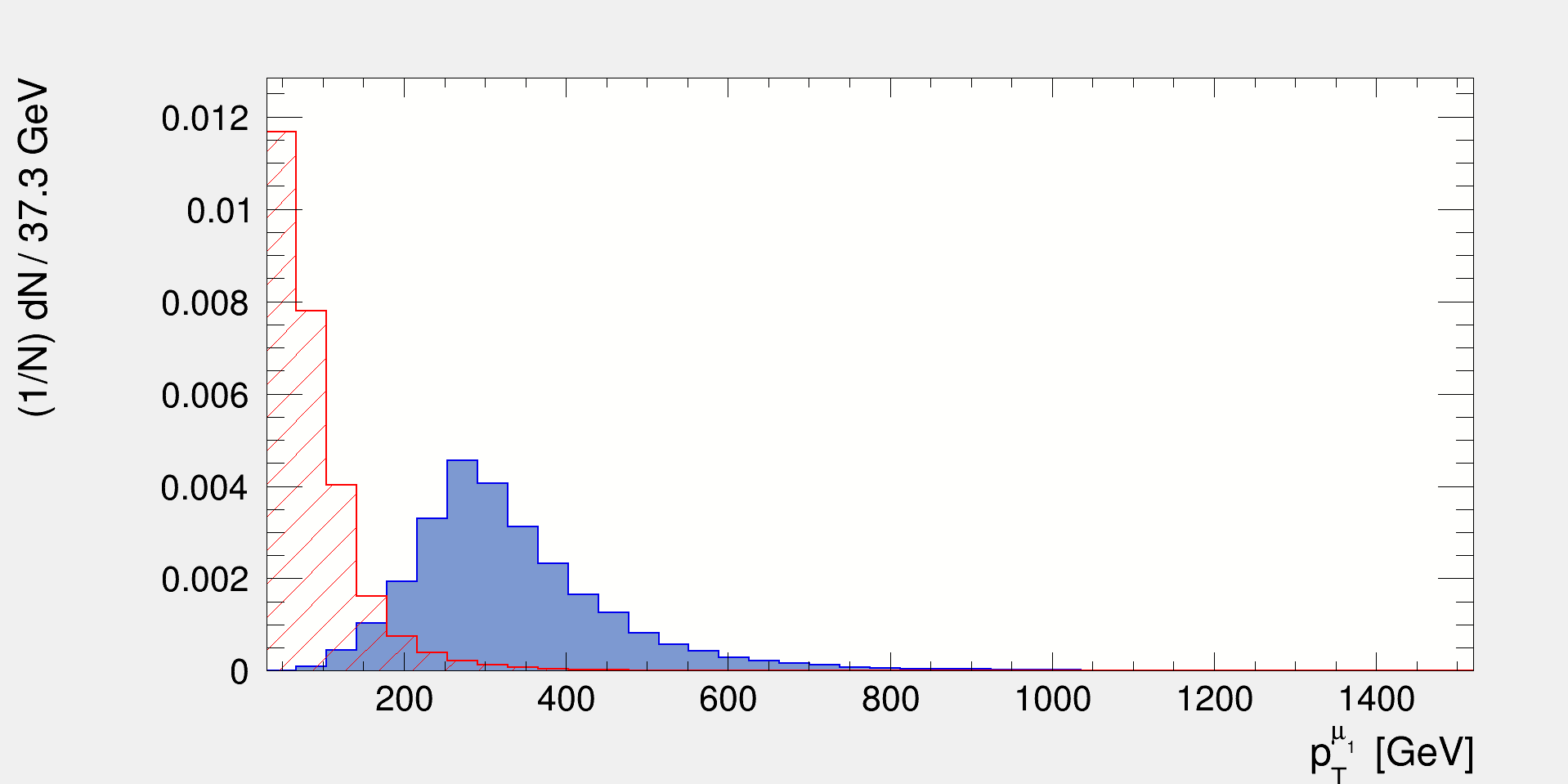}
    \caption{$P_T$ of $\mu_1$}
  \end{subfigure}
  \begin{subfigure}[b]{0.4\textwidth}
    \includegraphics[width=\textwidth]{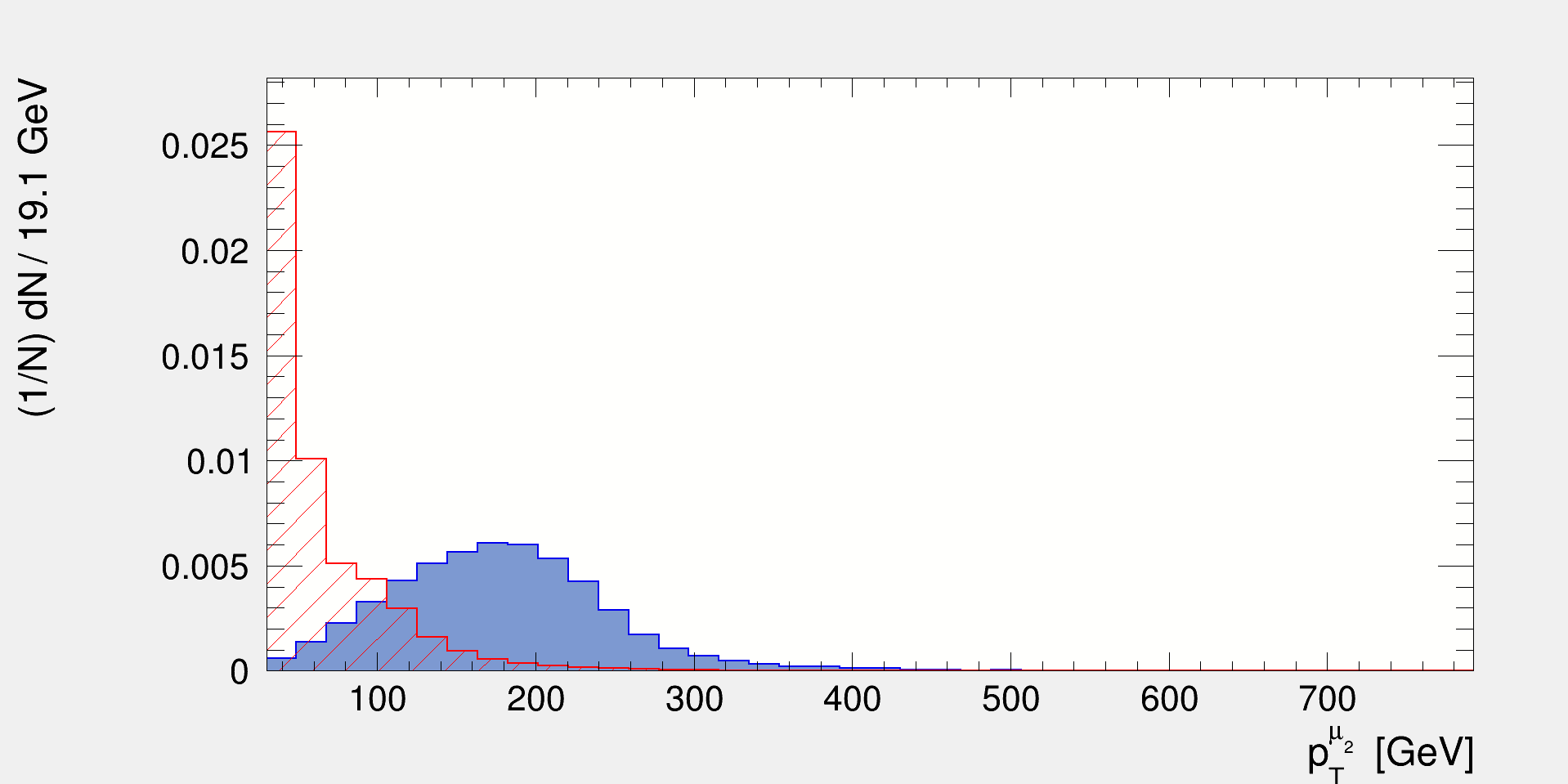}
    \caption{$P_T$ of $\mu_2$}
  \end{subfigure}
    \begin{subfigure}[b]{0.4\textwidth}
    \includegraphics[width=\textwidth]{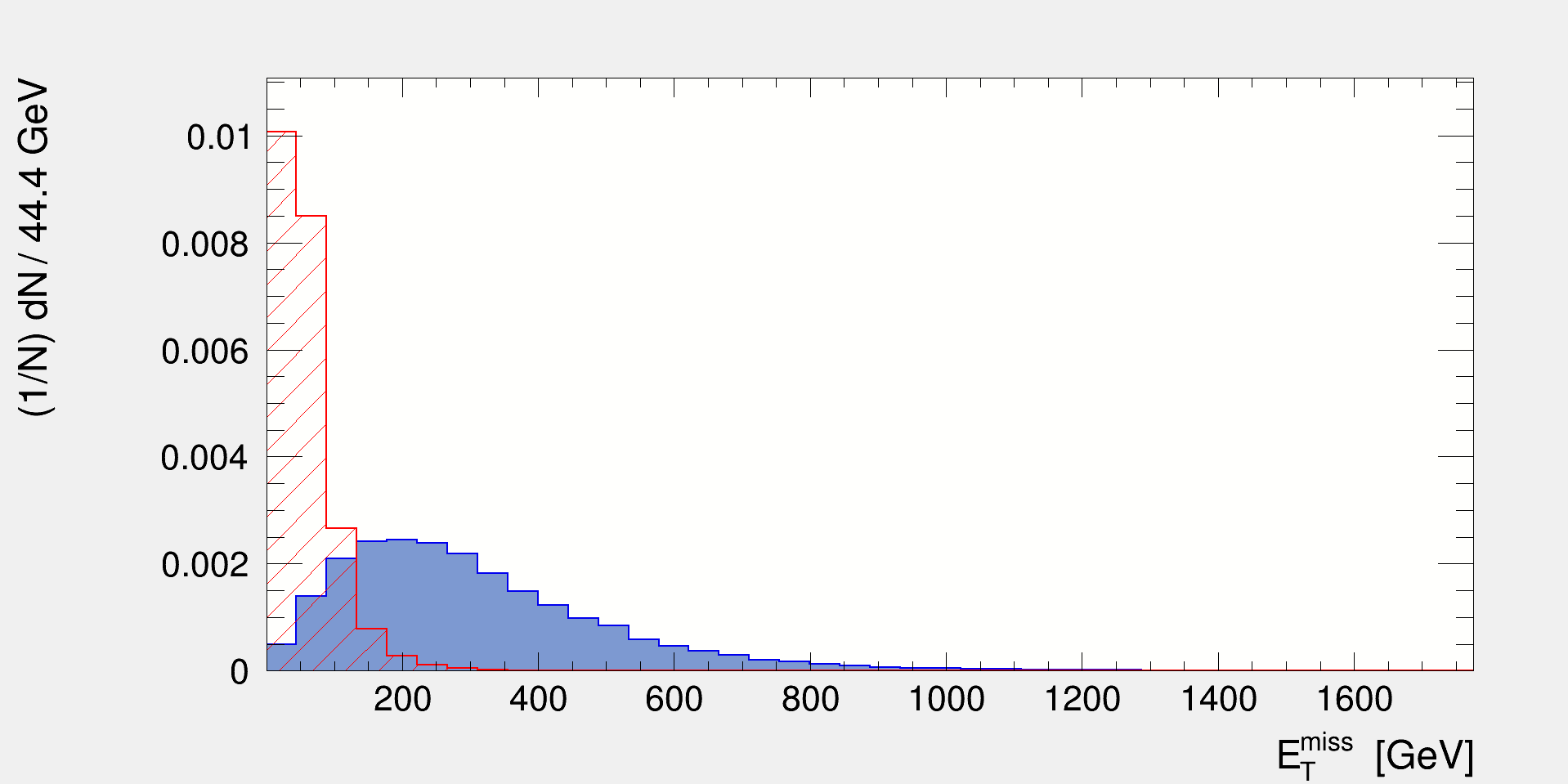}
    \caption{$E_{T}^{\mathrm{miss}}$}
  \end{subfigure}
      \begin{subfigure}[b]{0.4\textwidth}
    \includegraphics[width=\textwidth]{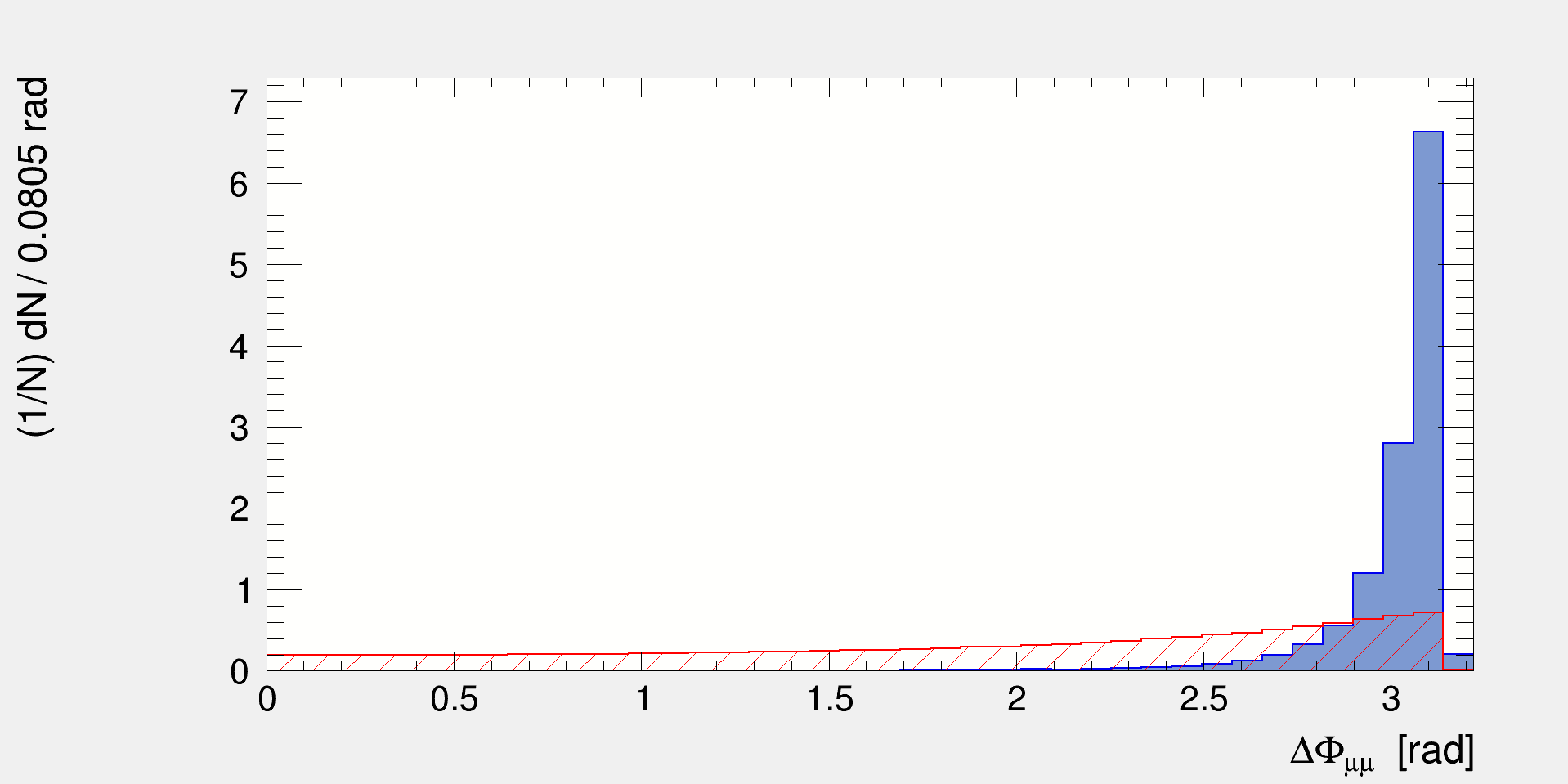}
    \caption{$\Delta\Phi$}
  \end{subfigure}

  \caption{Normalized distributions of kinematics used as features in the MVA classifiers for BP4 of the DH signal, taking $g_{SM} = 0.25$ and $g_{DM} = 1.0$, and its corresponding SM background, summarized in Table~\ref{tab:SM_backgrounds}, at $\sqrt{s} = 14,\mathrm{TeV}$.}
  \label{fig:Input_Vars_plots}
\end{figure*}
\begin{figure*}[htbp]
  \centering
  \begin{subfigure}[b]{0.45\textwidth}
    \includegraphics[width=\textwidth]{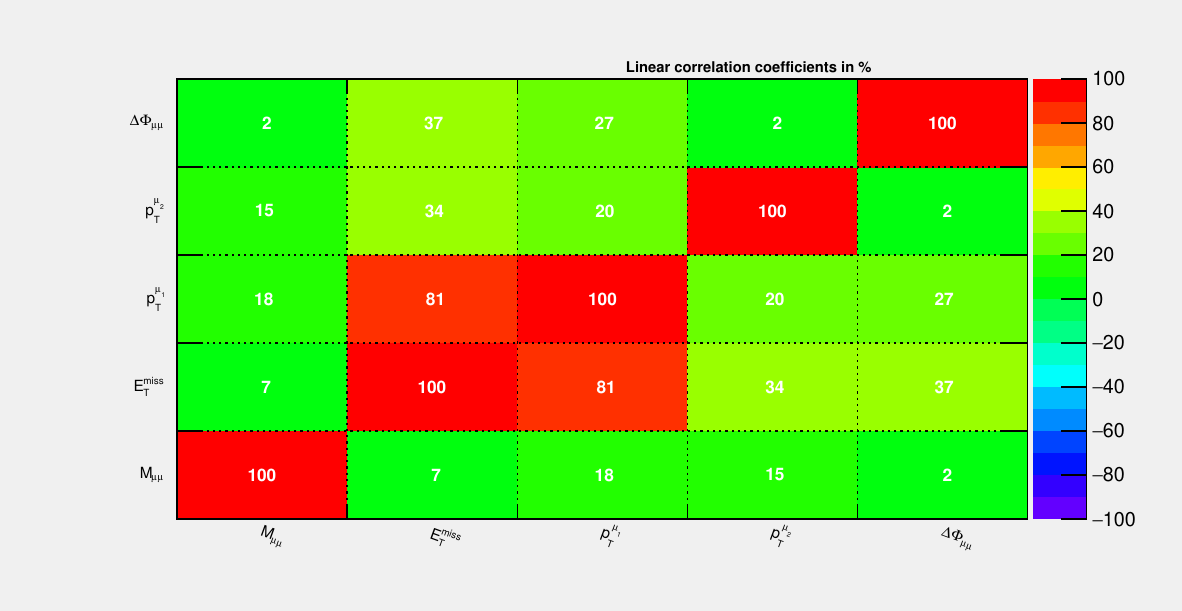}
    \caption{Signal}
  \end{subfigure}
  \begin{subfigure}[b]{0.45\textwidth}
    \includegraphics[width=\textwidth]{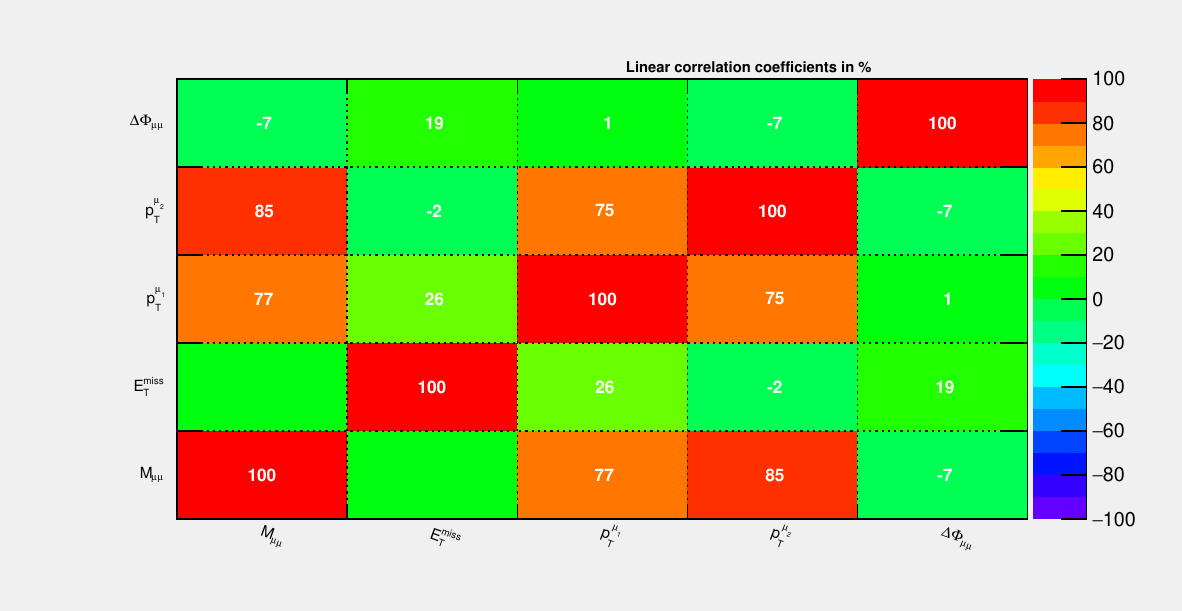}
    \caption{Background}
    \end{subfigure}
    
  \caption{Linear correlation matrix of the input features used in the MVA for (a) BP4 of the DH signal, taking $g_{SM} = 0.25$ and $g_{DM} = 1.0$, and (b) SM background, summarized in Table~\ref{tab:SM_backgrounds}, at $\sqrt{s} = 14,\mathrm{TeV}$. The linear correlation coefficients are displayed as percentages; the sign (positive or negative) indicates whether the variables are correlated or anti-correlated.}
  \label{fig:correlation}
\end{figure*}

Fig.~\ref{fig:Mass_Dist} displays the invariant mass distributions of the dimuon for different BPs of the DH signals overlaid on the overwhelming SM background. The dominant SM background is the DY, which is expected due to its high cross section listed in Table~\ref{tab:SM_backgrounds}. 

Examining Fig.~\ref{fig:Mass_Dist},  it becomes evident that distinguishing between signal and background is essential, motivating the use of MVA. Unlike cut-based analysis, which selects a single signal region, an MVA method, such as BDT, divides the phase space into numerous hypercubes, which are then classified by the BDT as either "background-like" or "signal-like", and a non-linear boundary is applied to discriminate the signal from the background effectively.

The MVA classifiers require a set of input variables for training, which should be linearly uncorrelated and descriptive to enable effective discrimination between signal and background. Table~\ref{tab:Input_Vars_table} lists the kinematics used for this task with their description.

\begin{figure}[htbp]
  \centering
  \includegraphics[width=0.4\textwidth]{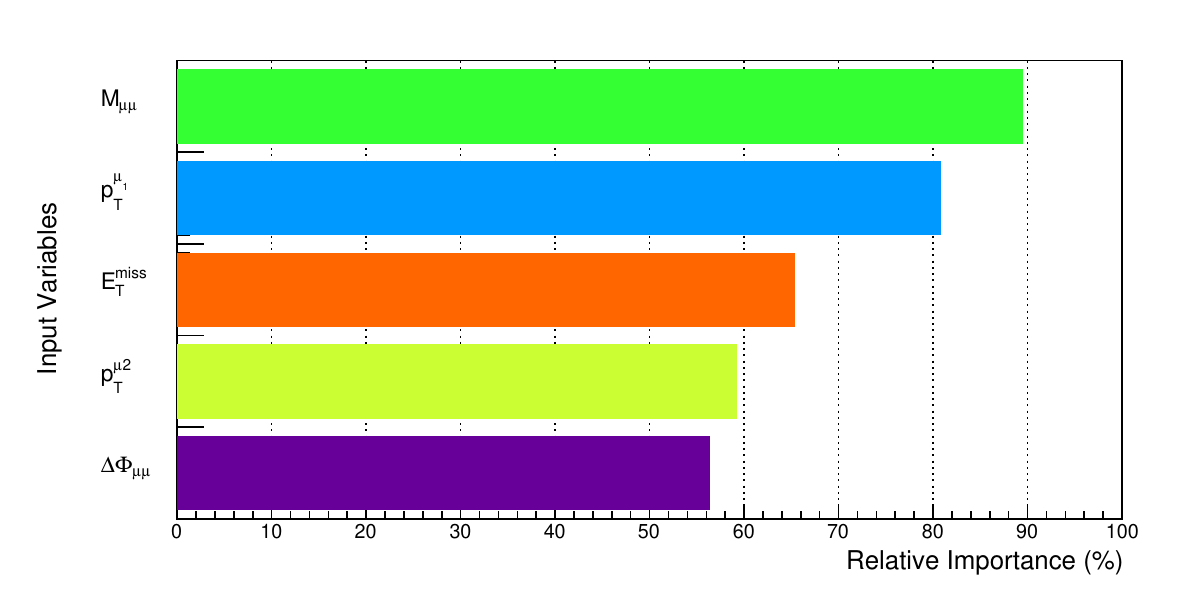} 
  \caption{The method-unspecific relative importance of the kinematics used in MVA for BP4 of the DH signal, taking $g_{SM} = 0.25$ and $g_{DM} = 1.0$, and its corresponding SM background, summarized in Table~\ref{tab:SM_backgrounds}, at $\sqrt{s} = 14,\mathrm{TeV}$. These numbers are obtained by the TMVA using Equation~\ref{Eq6.2}.}
  \label{fig:importance} 
\end{figure}

After thoroughly reviewing the distributions of the kinematics employed in this MVA, displayed in Fig.~\ref{fig:Input_Vars_plots}, the linear correlation coefficients $\zeta(C,D)$ between them are determined using 
Eq.~\ref{Eq6.1}~\cite{Bhardwaj2020Boosted}.
\begin{equation}
    \label{Eq6.1}
    \zeta(C,D) = \frac{F(CD) - F(C)F(D)}{\nu(C)\nu(D)}
\end{equation}
 where $F(CD)$, $F(C)$, and $F(D)$ are the expectation values for variables $C,\ D,\ \text{and}\ CD$, respectively, while $\nu(C)$ and $\nu(D)$ are the standard deviations of these variables. The importance of the linear correlation lies in determining how unique the information carried by the variable is. Fig.~\ref{fig:correlation} shows the correlation matrix for the input variables, showing that most of them are highly uncorrelated with each other. Certain variables exhibit a relatively higher degree of correlation in the signal samples compared to the background, or vice versa, such as ($P_{T}^{\mu_{1}}$, $P_{T}^{\mu_{2}}$, $M_{\mu\mu}$, and $E_T^{\mathrm{miss}}$). This is due to the differing kinematics between the signal and background processes. However, these variables are still used despite their high correlation because they are highly relevant, as indicated by Fig.~\ref{fig:importance}.

 The method-unspecific importance of the input variables, shown in Fig.~\ref{fig:importance}, is calculated according to their separation power $\Delta_\gamma$, ranging from 0 to 1, using Eq.~\ref{Eq6.2}~\cite{Bhardwaj2020Boosted}.
 \begin{equation}
     \label{Eq6.2}
     \Delta_\gamma = \int \frac{(\hat{P_s}(\gamma) - \hat{P_b}(\gamma))^2}{\hat{P_s}(\gamma) - \hat{P_b}(\gamma)}
 \end{equation}
 where $\gamma$ is the input variable required to find the separation power, and its probability distribution function (PDF) is denoted as $\hat{P_s}$ and  $\hat{P_b}$ for signal and background, respectively.
\begin{table*}[htbp]
\centering
\begin{tabular}{|p{3.7cm}|p{4cm}|p{6cm}|}
\hline
\multicolumn{3}{|c|}{\textbf{Likelihood}} \\
\hline
Parameter & Value & Description \\
\hline
TransformOutput & Enabled & Apply output transformation \\
PDFInterpol & Spline2 & PDF interpolation method \\
NSmoothSig[0] & 20 & Smoothing for signal PDF (var 0) \\
NSmoothBkg[0] & 20 & background Smoothing PDF (var 0) \\
NSmoothBkg[1] & 10 & background Smoothing PDF (var 1) \\
NSmooth & 1 & Global smoothing switch \\
NAvEvtPerBin & 50 & Avg. events per histogram bin \\
\hline
\multicolumn{3}{|c|}{\textbf{DNN}} \\
\hline
Layout &  \makecell[l]{TANH|128,\\ TANH|128,\\ TANH|128, LINEAR} & Network layers and activations \\
LearningRate & 1e-2 & Learning step size \\
Momentum & 0.9 & Gradient momentum \\
ConvergenceSteps & 20 & Early stopping patience \\
BatchSize & 100 & Samples per training batch \\
TestRepetitions & 1 & Test repetitions per epoch \\
WeightDecay & 1e-4 & L2 regularization strength \\
Regularization & None & Regularization type \\
DropConfig & 0.0 + 0.5 + 0.5 + 0.5 & Dropout rates per layer \\
\hline
\multicolumn{3}{|c|}{\textbf{BDT}} \\
\hline
NTrees & 1000 & Number of boosting trees \\
MinNodeSize & 2.0\% & Minimum leaf size \\
BoostType & AdaBoost & Boosting algorithm \\
AdaBoostBeta & 0.5 & Learning rate \\
UseBaggedBoost & Yes & Use bagging during training \\
BaggedSampleFraction & 0.5 & Fraction of data per bag \\
nCuts & 30 & Cuts per split search \\
MaxDepth & 3 & Max tree depth \\
SeparationType & GiniIndex & Node splitting metric \\
\hline
\end{tabular}
\caption{List of hyperparameters for the BDT, DNN, and Likelihood methods, along with their descriptions and corresponding values used in this MVA.}
\label{tab:hyperparams}
\end{table*}
\begin{figure*}[htbp]
  \centering
  \begin{subfigure}[b]{0.4\textwidth}
    \includegraphics[width=\textwidth]{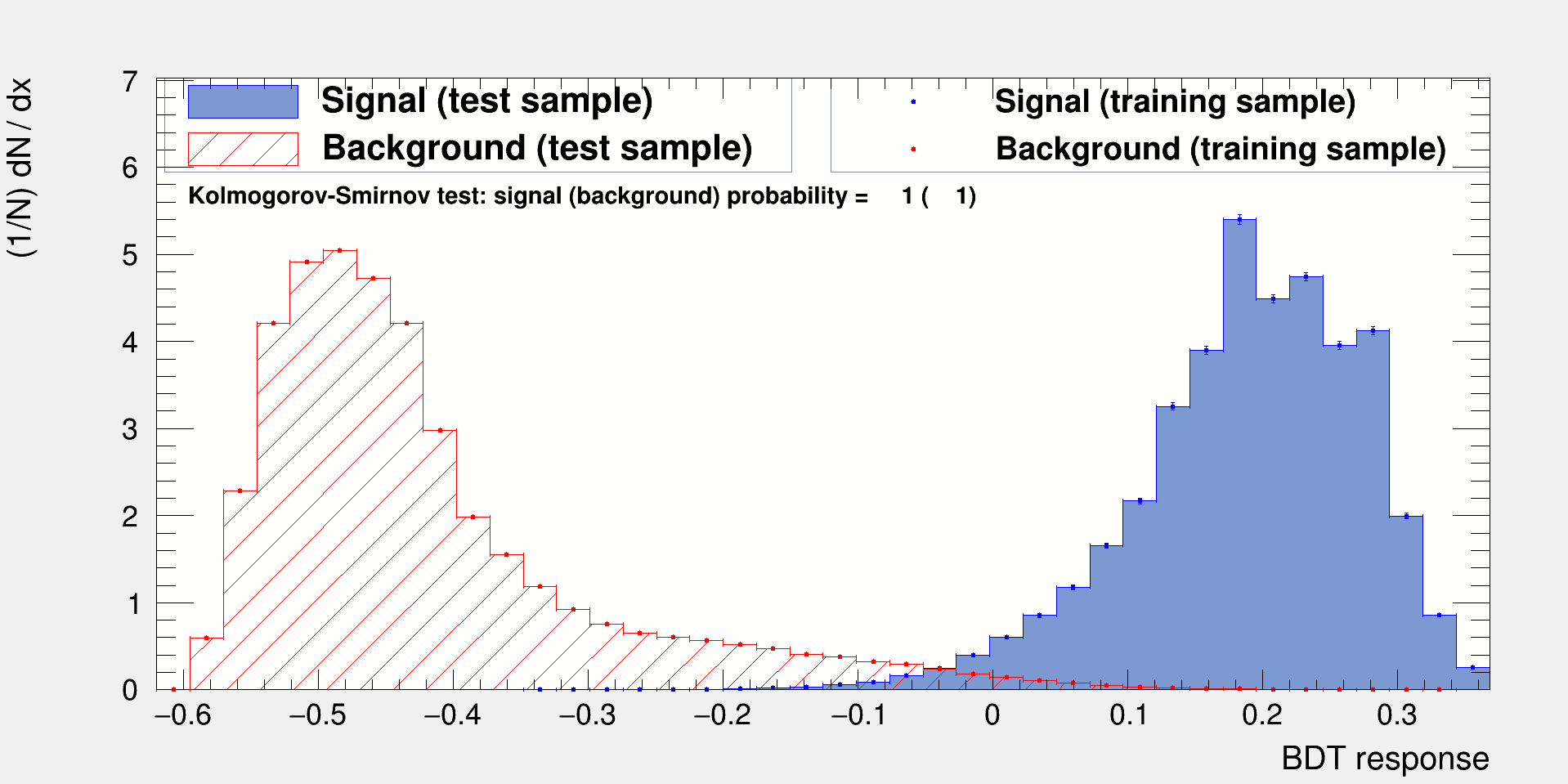}
    \caption{BDT}
  \end{subfigure}
  \begin{subfigure}[b]{0.4\textwidth}
    \includegraphics[width=\textwidth]{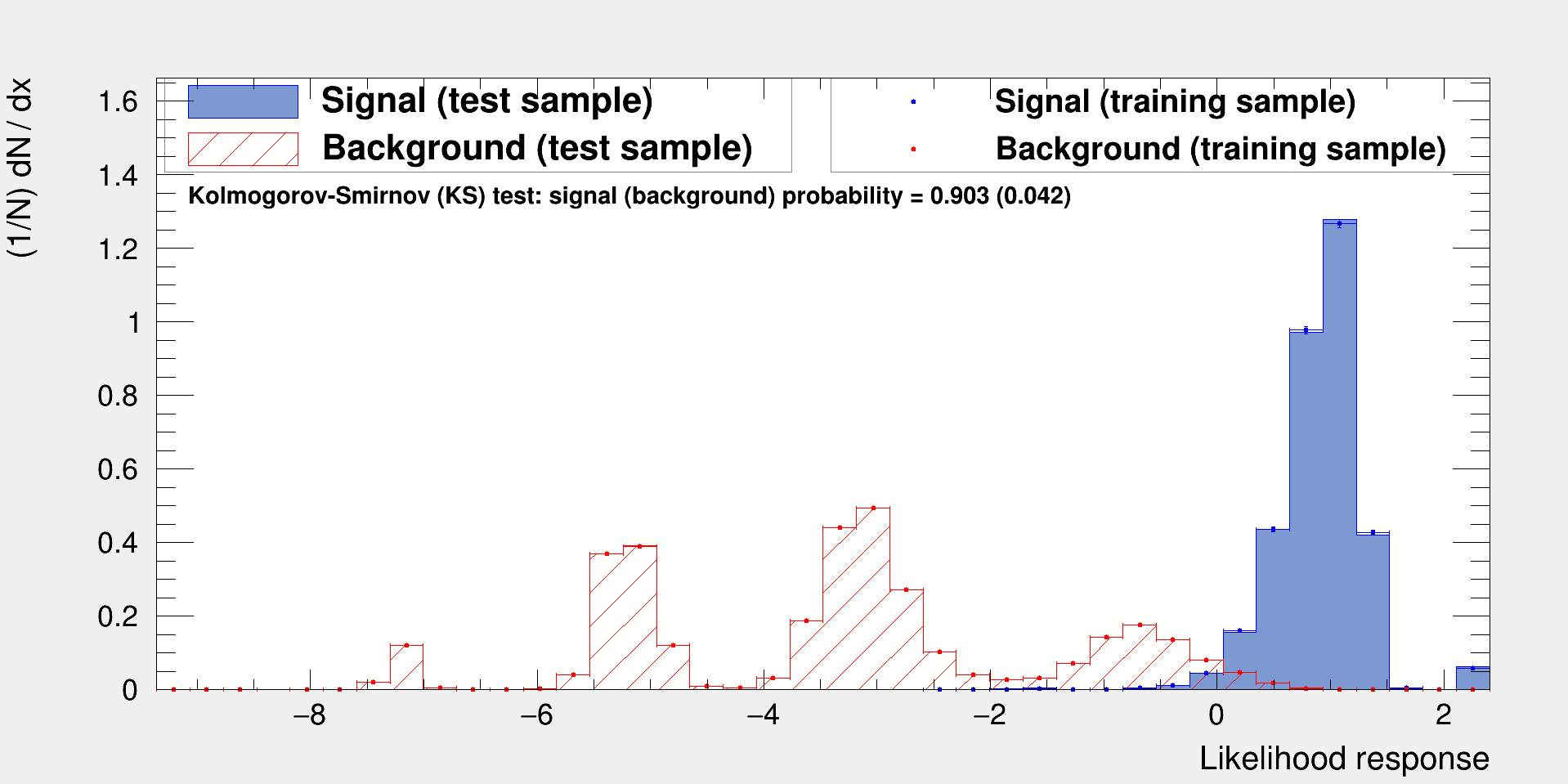}
    \caption{Likelihood}
    \end{subfigure}
  \begin{subfigure}[b]{0.4\textwidth}
    \includegraphics[width=\textwidth]{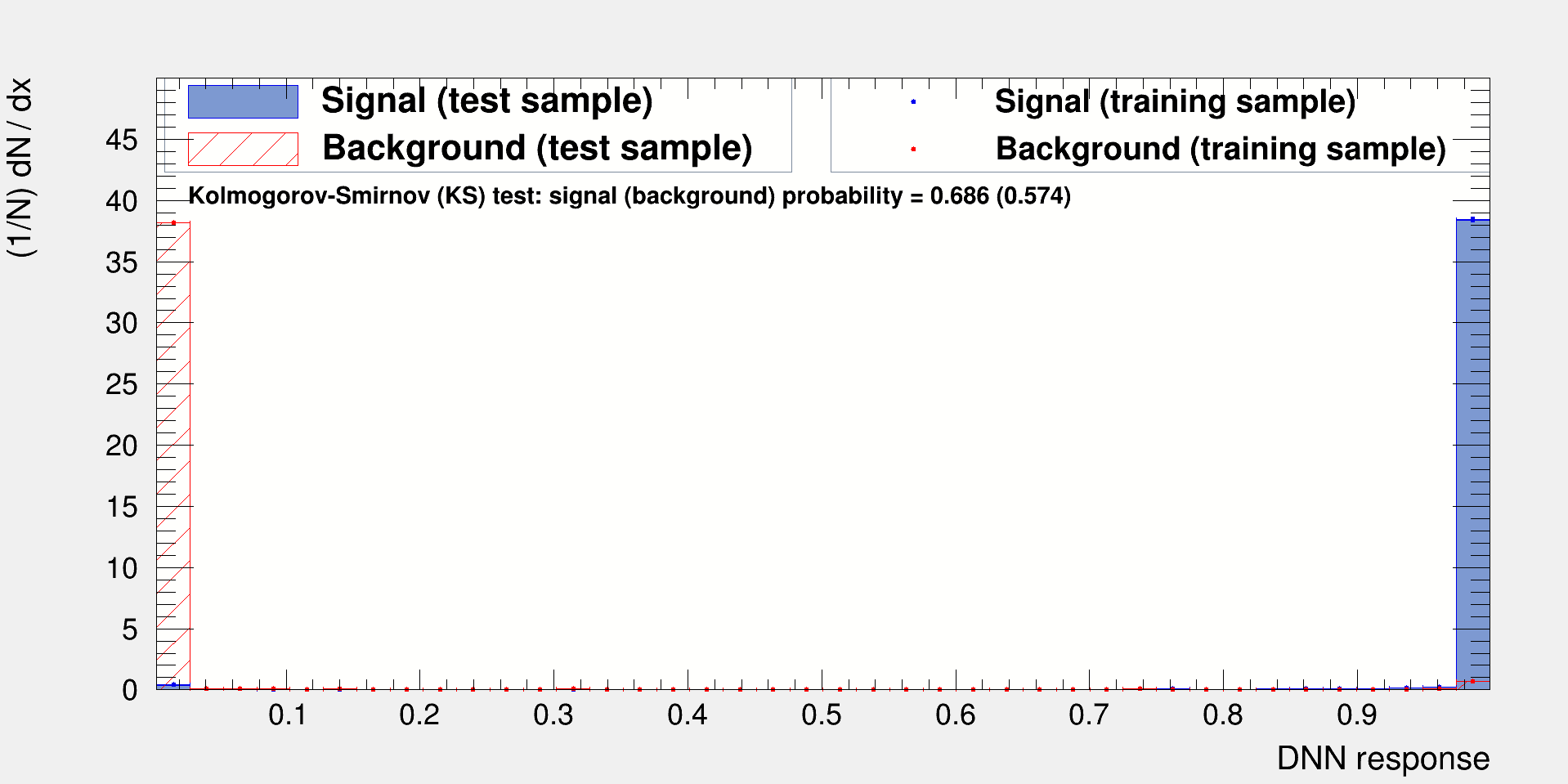}
    \caption{DNN}
    \end{subfigure}
    
  \caption{Classifier response distributions for training and testing samples of (a) BDT, (b) Likelihood, and (c) DNN methods for BP4 of the DH signal, taking $g_{SM} = 0.25$ and $g_{DM} = 1.0$, along with the corresponding SM background, summarized in Table~\ref{tab:SM_backgrounds}, at $\sqrt{s} = 14,\mathrm{TeV}$.}
  \label{fig:Response}
\end{figure*}

The classifiers are trained using the $k$-fold cross-validation (CV) approach, with $k=5$, for each BP of the DH signal and its corresponding SM background. In this method, the dataset is divided into $k$ equal parts (folds). The model is trained on $k - 1$ of these folds while the remaining fold is used for validation. This procedure is iterated $k$ times, with each fold serving as the validation set exactly once. The CV approach provides a robust estimate of model performance and helps mitigate overfitting, especially in HEP analyses where data imbalance or statistical fluctuations are common~\cite{Kohavi1995,Arlot2010,Refaeilzadeh2009}. The hyperparameters used for the classifiers, BDT, DNN, and Likelihood, are summarized in Table~\ref{tab:hyperparams}.

The overtraining check plots for the three classifiers are shown in Fig.~\ref{fig:Response}, demonstrating that the testing data closely matches the training data. This indicates good training performance for all three classifiers, as expected due to the use of CV. To quantitatively assess potential overtraining, the Kolmogorov–Smirnov (KS) test is employed. The KS test compares the cumulative distributions of the training and testing samples for each classifier's output, providing a statistical measure of their agreement~\cite{Massey1951}. A high p-value from the KS test indicates that the two distributions are statistically similar, confirming that the classifiers generalize well to unseen data.

\begin{figure}[htbp]
  \centering
  \includegraphics[width=0.4\textwidth]{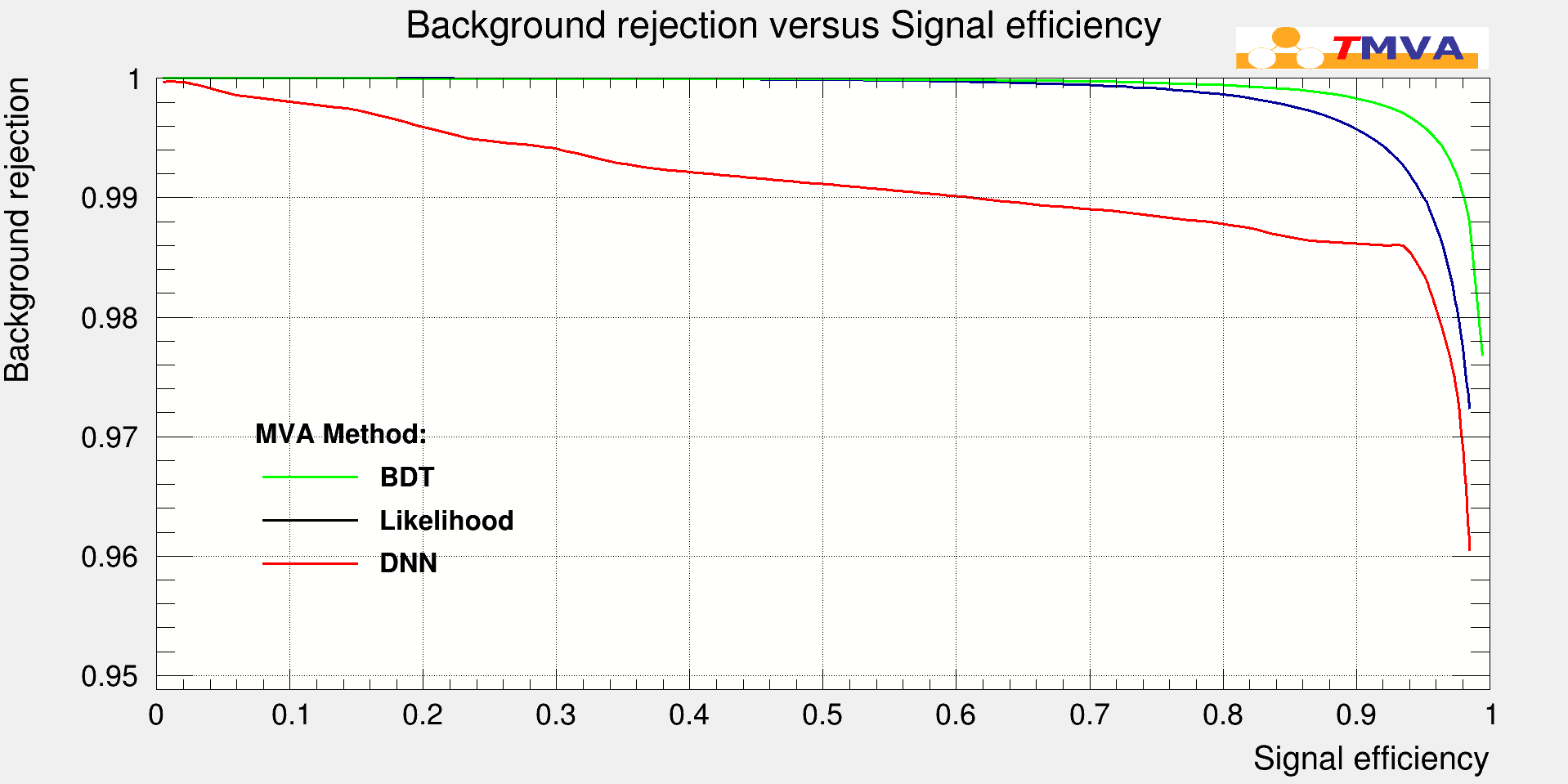} 
  \caption{ROC curves comparing the performance of the BDT, DNN, and Likelihood classifiers used in this MVA study. The BDT method exhibits the highest AUC, followed by the Likelihood classifier and then the DNN. A higher AUC indicates better discriminating power between signal and background.}
  \label{fig:Roc} 
\end{figure}

Fig.~\ref{fig:Roc} shows the ROC curves for the three classifiers, plotting background rejection versus signal efficiency. The area under the curve (AUC) serves as a metric for classifier performance, with values ranging from 0 to 1. A higher AUC indicates better discriminative power. Based on the AUC values in Fig.~\ref{fig:Roc}, the best classifier is the BDT, followed by the likelihood method, with the DNN performing the worst.

Thus, the study proceeds with the BDT classifier to calculate the statistical significance ($z$) for all BPs of the DH signal against the corresponding SM background.

\begin{figure}[htbp]
  \centering
  \includegraphics[width=0.4\textwidth]{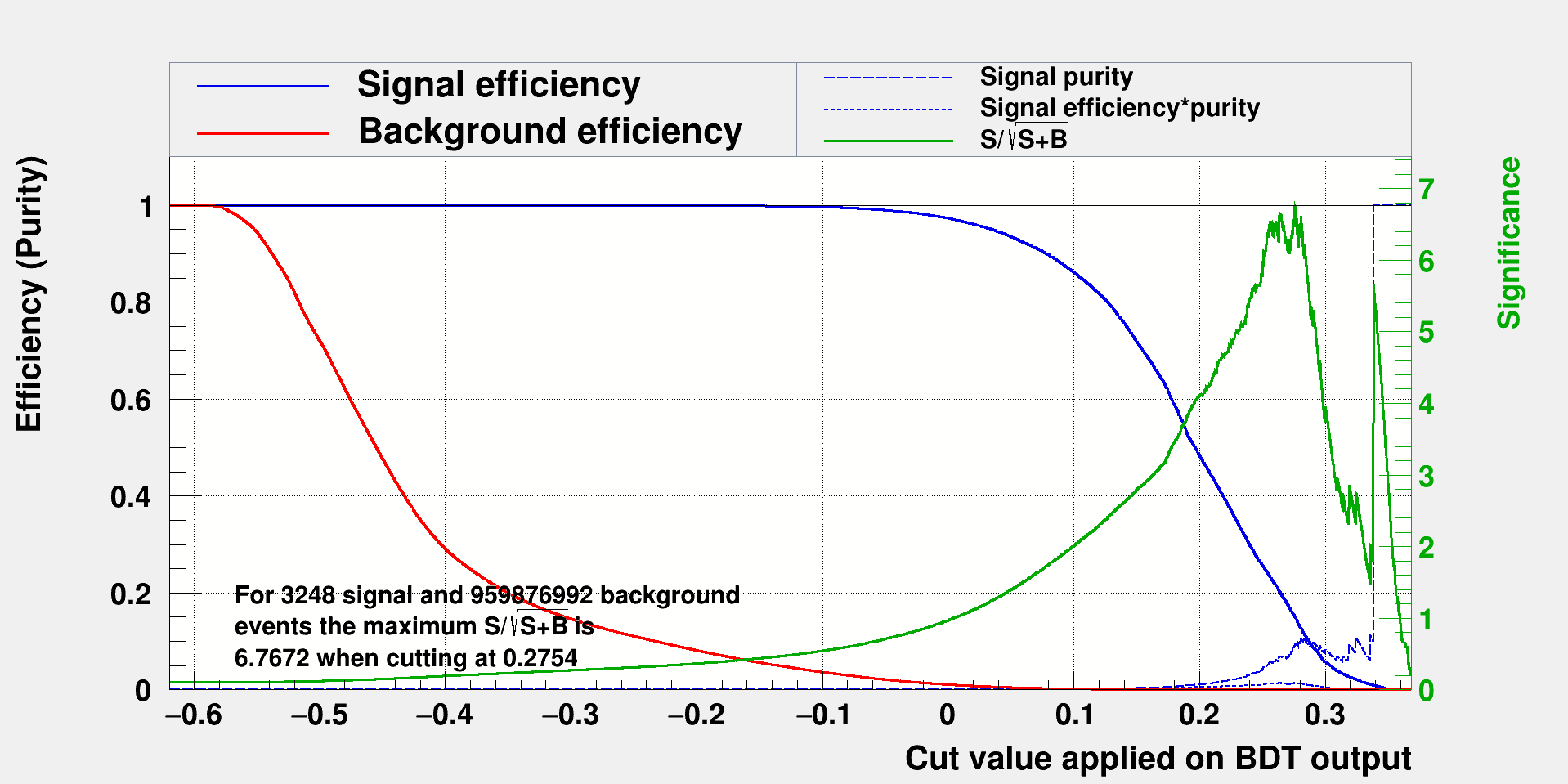} 
  \caption{Cut efficiencies verses BDT response thresholds evaluated for BP4 of the DH signal, taking $g_{SM} = 0.25$ and $g_{DM} = 1.0$, along with the corresponding SM background, summarized in Table~\ref{tab:SM_backgrounds}, with $\mathcal{L} = 3000\,\mathrm{fb}^{-1}$ at $\sqrt{s} = 14\,\mathrm{TeV}$.}
  \label{fig:Effeciency}
\end{figure}

The required luminosity to achieve a 5$\sigma$ significance ($\mathcal{L}_{5\sigma}^{\text{Req}}$ [fb$^{-1}$]) over the SM background for the first four BPs of the DH signal at $\sqrt{s} = 14,\mathrm{TeV}$ is shown in Table~\ref{tab:5_sigma}. Since 5$\sigma$ significance is achieved for the first three BPs at $\mathcal{L} < 500$~fb$^{-1}$, their z values are not calculated at $\mathcal{L} = 3000$~fb$^{-1}$, as the resulting values would be exceedingly large. Consequently, a cut value is applied to the BDT response for BP4 through BP9 to obtain the corresponding numbers of signal events ($S$) and background events ($B$) with $\mathcal{L} = 3000$~fb$^{-1}$ at $\sqrt{s} = 14,\text{TeV}$.  Finally, $z$ is calculated for these six BPs using Eq.~\ref{Eq6.3}~\cite{CMS:2021}.
\begin{equation}
    \label{Eq6.3}
    z = \frac{S}{\sqrt{S + B}}
\end{equation}
The applied cut value, also known as optimal cut, is where the maximum \textbf{z} is achieved for the BP as shown in Fig.~\ref{fig:Effeciency} for BP4.
\begin{table}[htbp]
\centering
\begin{tabular}{|c|c|}
\hline
\textbf{BPs} & $\bm{\mathcal{L}_{5\sigma}^{\text{Req}}\ [\mathrm{fb}^{-1}]}$ \\
\hline
BP1 & 4.3 \\
BP2 & 18.8 \\
BP3 & 175 \\
BP4 & 1650 \\
\hline
\end{tabular}
\caption{The Required luminosity to achieve 5$\sigma$ significance $\mathcal{L}_{5\sigma}^{\text{Req}}$ over the SM background, summarized in Table~\ref{tab:SM_backgrounds}, for the first four BPs of the DH signal, taking $g_{DM} = 1.0\ \text{and}\ g_{SM} = 0.25$, at $\sqrt{s} = 14,\mathrm{TeV}$.}
\label{tab:5_sigma}
\end{table}

Table~\ref{tab:full_tmva_results} presents results of $z$ for BP4 through BP9. It is evident that the BDT classifier achieves a high \textbf{z} for BP4 with $\mathcal{L} = 3000\,\mathrm{fb}^{-1}$ at $\sqrt{s} = 14\,\mathrm{TeV}$, indicating strong discovery potential for this mass configuration and lighter ones. In contrast, the other BPs, which correspond to higher $M_{Z^\prime}$ than that of BP4, do not reach the discovery threshold at the HL-LHC. This suggests that either increased $\mathcal{L}$ or a higher $\sqrt{s}$ may be required to effectively probe these higher-mass BPs.

\begin{table*}[htbp]
\centering
\small
\begin{tabular}{| c | c | c | c | c | c |}
\hline
\textbf{BPs} & $\bm{N^{BC}}$ & \textbf{Opt. Cut} & $\bm{S\ (\epsilon_S)}$ & $\bm{B\ (\epsilon_B)}$ & \textbf{z} \\
\hline
BP4 & 3248 & 0.2754 & 500 (15.38\%) & 4951 ($5.158\times10^{-6}$) & 6.77 \\
\hline
BP5 & 1471 & 0.2810 & 51 (3.53\%) & 290 ($3.021\times10^{-7}$) & 2.81 \\
\hline
BP6 & 716 & 0.2670 & 35 (4.87\%) & 583 ($6.069\times10^{-7}$) & 1.40 \\
\hline
BP7 & 372 & 0.2678 & 22 (5.95\%) & 291 ($3.034\times10^{-7}$) & 1.25 \\
\hline
BP8 & 203 & 0.2932 & 4 (1.82\%) & 27 ($2.798\times10^{-8}$) & 0.72 \\
\hline
BP9 & 114 & 0.2739 & 5 (4.47\%) & 121.16 ($1.262\times10^{-7}$) & 0.48 \\
\hline
\multicolumn{1}{|c|}{SM Bkg} & $9.59877 \times 10^8$ & - & - & - & - \\
\hline
\end{tabular}

\caption{Summary of BDT optimal cuts, events numbers before the cut $N^{BC}$, and after the cut for signal $S$, and backgrounds $B$. Also, listed the statistical significance $z$ for all BPs of the DH signal, taking $g_{DM} = 1.0\ \text{and}\ g_{SM} = 0.25$, along with the corresponding SM background, summarized in Table~\ref{tab:SM_backgrounds}, with $\mathcal{L} = 3000\,\mathrm{fb}^{-1}$ at $\sqrt{s} = 14\,\mathrm{TeV}$. Signal acceptance ($\epsilon_S$) and background acceptance ($\epsilon_B$) are shown in parentheses.}
\label{tab:full_tmva_results}
\end{table*}
\section{Summary}
\label{section:Summary}

In this study, a multivariate analysis (MVA) was carried out to investigate the discrimination between the Dark Higgs (DH) signal and the Standard Model (SM) backgrounds in the leptonic dimuon decay channel of the $Z'$ boson, within the context of proton-proton collisions at the High-Luminosity Large Hadron Collider (HL-LHC). 

Events were generated by taking $g_{SM} = 0.25$ and $g_{DM} = 1.0$ for a luminosity of $\mathcal{L} = 3000\,\mathrm{fb}^{-1}$ at a center-of-mass energy of $\sqrt{s} = 14\,\mathrm{TeV}$. A set of optimized kinematic variables was used as input variables for the MVA.

Three classifiers, BDT, DNN, and Likelihood, were implemented using TMVA. The BDT exhibited the highest performance according to the ROC integrals. The required luminosities to achieve 5$\sigma$ significance ($\mathcal{L}_{5\sigma}^{\text{Req}}$) were determined for the first four BPs. For BP1, BP2, and BP3, discovery can be achieved with $\mathcal{L} < 500~\text{fb}^{-1}$, and therefore their significances z at $\mathcal{L} = 3000~\text{fb}^{-1}$ were not evaluated, as they would be exceedingly large. For BP4 through BP9, an optimal cut on the BDT response was applied to maximize the significance.

The results show that the BDT achieves high z values at $M_{Z'} = 500~\text{GeV}$ (BP4) under HL-LHC conditions. However, for the higher-mass BPs (i.e., BP5 through BP9), discovery remains unachievable even at $\mathcal{L} = 3000~\text{fb}^{-1}$. This highlights the sensitivity limits for heavy-mass $Z'$ searches in the DH model at the HL-LHC. Future studies incorporating systematic uncertainties and additional channels may extend the discovery potential.
%

\end{document}